\documentclass[12pt]{article}

\usepackage{graphicx}        
\usepackage{multicol}        
\usepackage[bottom]{footmisc}

\def\be{\begin{equation}}
\def\ee{\end{equation}}
\def\beq{\begin{eqnarray}}
\def\eeq{\end{eqnarray}}

\def\m{\mu}
\def\n{\nu}

\begin{document}

\begin{titlepage}

\begin{centering}

{\huge {\bf Kac-Moody algebras and the structure of cosmological singularities: a new light on the Belinskii-Khalatnikov-Lifshitz analysis}}\footnote{Dedicated to Claudio Bunster on the occasion of his 60th birthday. To appear in ``Quantum Mechanics of Fundamental Systems: The Quest for Beauty and Simplicity - Claudio Bunster Festsschrift"}

\vspace{1cm}

{\Large Marc Henneaux$^{a,b}$} \\
\vspace{.7cm}
$^a$ Physique Th\'eorique et Math\'ematique,
Universit\'e Libre
de Bruxelles  and
International Solvay Institutes, \\ ULB C.P. 231, B-1050, Bruxelles, Belgium      \\
\vspace{.2cm}
$^b$ Centro de Estudios Cient\'{\i}ficos, Casilla 1469, Valdivia, Chile \\

\vspace{1.5cm}

\end{centering}

\begin{abstract}
The unexpected and fascinating emergence of hyperbolic Coxeter groups and Lorentzian Kac-Moody algebras in the investigation of gravitational theories in the vicinity of a cosmological singularity is briefly reviewed. Some open questions raised by this intriguing result, and some attempts to answer them, are outlined.
\end{abstract}

\vfill
\end{titlepage}

\section{Introduction}
\label{sec:1}
As reported by Khalatnivov \cite{Khalatnikov:2008zt}, the problem of cosmological singularities was considered by Lev Landau as one of the three most important problems of theoretical physics.  While great breakthroughs occurred in the understanding of superconductivity and phase transitions -- the other two important problems identified by Landau --, taming singularities in gravity theory (understanding their structure and their possible resolution in an appropriate completion of gravity) remains to this day a challenge which raises a vast number of unanswered questions, in spite of the important advances achieved in the field in the last 50 years or so.

A major development in the area was the construction by Belinskii, Khalatnikov and Lifschitz (``BKL") \cite{Belinsky:1970ew,Belinsky:1982pk} of a general solution of the gravitational field equations in the vicinity of a spacelike (``cosmological") singularity. Their work led recently to new investigations pointing to a fascinating and somewhat unexpected connection between gravity and Lorentzian Kac-Moody algebras \cite{DH2,DHNRev}.  The purpose of article is to briefly review these recent developments and to provide a guide to the literature on the subject.

It is a pleasure to dedicate this article to Claudio Bunster, long term collaborator and friend, on the occasion of his 60th birthday.  The choice of cosmological singularities is a particularly appropriate subject since it is Claudio (then named ``Claudio Teitelboim") who introduced some 30 years ago the author to the remarkable BKL analysis (interest in the BKL analysis was then motivated by the desire to understand the ``zero signature" limit of gravity \cite{Teitelboim:1978uc,Henneaux:1979vn,Teitelboim:1981fb,Henneaux:1981su}).

\section{Original BKL Analysis and Extension to Higher Dimensions}

In their investigation of the generic dynamical behavior of the gravitational field in the vicinity of a cosmological (= spacelike) singularity, Belinskii, Khalatnikov and Lifschitz discovered the following remarkable features in four dimensions \cite{Belinsky:1970ew,Belinsky:1982pk}:
\begin{itemize}
\item As one reaches the singularity, the spatial points decouple in the sense that the dynamical Einstein equations, which are partial differential equations, become ordinary differential equations with respect to time (one finite number of ODE's at each spatial point).
\item In that limit, the off-diagonal components of the metric freeze (i.e., tend to definite limits) so that the non trivial dynamics is carried by the 3 independfent scale factors that indicate how distances along 3 independent spatial directions change with time\footnote{See \cite{Belinsky:1970ew,Belinsky:1982pk,DHNRev} for more information on the choice of slicing adapted to the singularity and the definition of scale factors.}.
\item The dynamics of the scale factors exhibit a never-ending, oscillatory behavior of chaotic type with an infinite number of oscillations as one goes to the singularity (see also \cite{Misner}).
\end{itemize}
This work was reformulated by Chitre and Misner in terms of a billiard motion in the 2-dimensional hyperbolic space of the dynamically independent scale factors (the 3 scale factors are related by the Hamiltonian constraint) \cite{Chitre,Misner2}.  Chaos is related in that picture to the finiteness of the volume of the billiard table. This reformulation turns out to be crucial for exhibiting the symmetries.

The extension to higher dimensions was started by Belinskii and Khalatnikov in 5 dimensions \cite{Belinsky:1988mc} and continued in \cite{Demaret:1986su,Demaret:1986ys,Demaret:1989wh} to all spacetime dimensions, with the surprising result that while the first two features (decoupling of spatial points and freezing of off-diagonal components) still hold, chaos disappear in spacetime dimensions $\geq 11$. The infinite number of oscillations of the scale factors is replaced asymptotically by a monotonic Kasner regime.  In particular, $11$-dimensional pure gravity is non chaotic. However, if one includes the $3$-form of $11$-dimensional supergravity, chaos reappears \cite{DamourHenneaux1}.

The understanding of the higher dimensional dynamics in terms of a billiard motion was also achieved and led to the same picture of a cosmological billiard ball moving in an hyperbolic space of higher dimension \cite{Russians,Russians2,DH2,DHNRev}.  This picture still holds if one includes dilatons and $p$-forms: the dilatons play the role of extra scale factors, while the $p$-form components play the role of extra off-diagonal components.

It should be stressed that the emergence of chaos for those models that are chaotic is a statement valid for generic initial data.  Chaos may be absent in models with particular spacetime symmetries, which form a set of measure zero.  This corresponds to removing billiard walls and enlarging thereby the billiard table, making its originally finite volume infinite (see \cite{Demaret:1988sg,Damour:2000th}).

\section{Emergence of Coxeter Groups and Kac-Moody Algebras}

The billiard picture just described holds for any Lagrangian of the form
\beq &&S[g_{\m \n}, \phi,
A^{(p)}] = \int d^D x \, \sqrt{- \, \!^{(D)}g} \;
\Bigg[R - \sum_i\partial_\m \phi_i \partial^\m \phi_i \nonumber \\
&& \hspace{1.5cm} - \frac{1}{2} \sum_p \frac{1}{(p+1)!}
e^{\l^{(p)} \phi} F^{(p)}_{\m_1 \cdots \m_{p+1}} F^{(p)  \, \m_1
\cdots \m_{p+1}} \Bigg] + \hbox{``more"} \label{keyaction} \eeq
where $D$ is the spacetime dimension.  The integer $p \geq 0$ labels the various
$p$-forms $A^{(p)}$ present in the theory, with field strengths
$F^{(p)}$ equal to $dA^{(p)}$, \be F^{(p)}_{\m_1 \cdots \m_{p+1}}
= \partial_{\m_1} A^{(p)}_{\m_2 \cdots \m_{p+1}} \pm p \hbox{
permutations }.  \ee In fact, the field strength could be modified
by additional coupling terms of Yang-Mills or Chapline-Manton type
\cite{pvnetal,CM} (e.g., $F_C = dC^{(2)} - C^{(0)} dB^{(2)}$ for
two $2$-forms $C^{(2)}$ and $B^{(2)}$ and a $0$-form $C^{(0)}$, as
it occurs in ten-dimensional type IIB supergravity), but we
include these additional contributions to the action in  ``more".
Similarly, ``more" might contain Chern-Simons terms, as in the
action for eleven-dimensional supergravity \cite{CJS}. The real parameters $\l^{(p)_i}$ measure the strengths of
the couplings to the dilatons.

However, a new feature emerges for theories which have the property that when reduced to three dimensions on a torus, their Lagrangian equals (after dualization to scalars of all the fields that can be dualized) the sum of the standard Einstein-Hilbert action plus the scalar Lagrangian of the non-linear sigma-model $G/H$, where $G$ is some simple Lie group and $H$ its maximal compact subgroup,
\begin{equation} \mathcal{L} = \mathcal{L}_E + \mathcal{L}_{G/H} . \end{equation} This class of theories include pure gravity in $D= d+1$ dimensions (for which the group $G$ is $SL(d-1,R)$ and $H = SO(d-1)$ \cite{Breitenlohner:1987dg,Cremmer:1999du}), or eleven-dimensional supergravity, for which $G = E_{8,8}$ and $H= SO(16)$ \cite{Cremmer:1979up,Marcus:1983hb}.

The crucial feature that emerges in that case is that the billiard table is then a Coxeter polyhedron and hence the billiard group (generated by the reflections in the billiard walls) is a Coxeter group \cite{DH2}.  This means that the angles of the billiard table are acute and equal to integer submultiples of $\pi$ (see \cite{Henneaux:2007ej} for information on Coxeter groups relevant to this context).  Furthermore, the Coxeter polyhedron is a simplex and the matrix built out of the scalar products of the wall forms $w_i$ defining the billiard
\begin{equation} A_{ij} = 2 \frac{(w_i \vert w_j)}{(w_i \vert w_i)}
\end{equation}
turns out to be the Cartan matrix of the Lorentzian Kac-Moody algebra $G^{++}$ \cite{DH2,deBuyl}. Here, $G^{++}$ denotes the overextension \cite{Julia,FF,Kac} of the algebra $G$.  Namely, it is obtained by adding a further simple root to the untwisted affine extension $G^+$ of $G$.  That root, the ``overextended root" is attached to the affine root by a single line. We give here a few examples.
\begin{itemize}
\item The algebra $A_1^{++}$ relevant to pure, four-dimensional gravity.
\begin{figure}[h]
\begin{center}
\includegraphics[scale=.6]{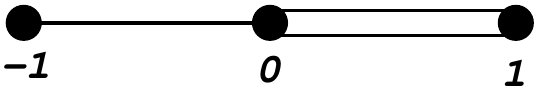}
\end{center}
\caption{The hyperbolic algebra $A_1^{++}$. The node $1$ defines the Dynkin diagram of $A_1$, the nodes $1$ and $0$ form the Dynkin diagram of its affine extension, while the nodes $1$, $0$ and $-1$ define its overextension $A_1^{++}$.}
\end{figure}

\item The algebra $E_8^{++} \equiv E_ {10}$ relevant to eleven-dimensional supergravity
\begin{figure}[h]
\begin{center}
\includegraphics[scale=.6]{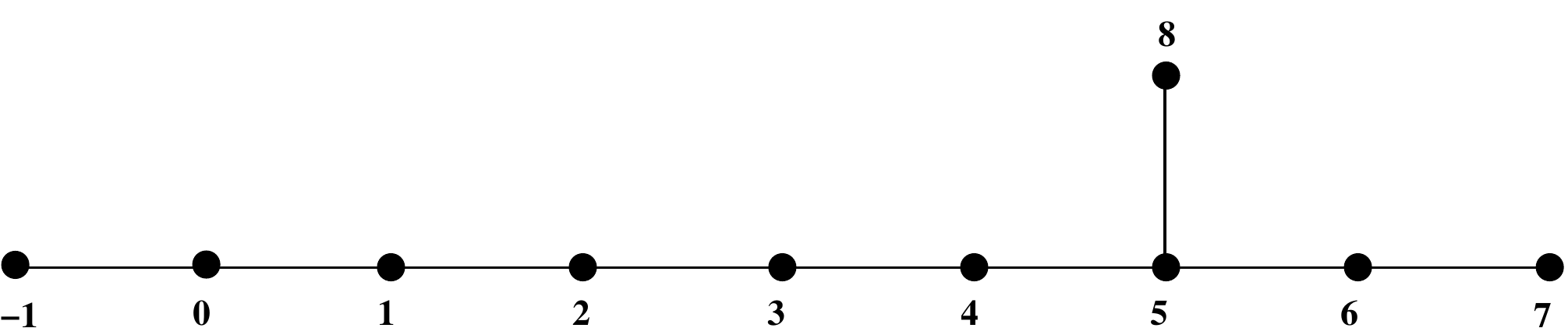}
\end{center}
\caption{The hyperbolic algebra $E_{10}$. The nodes labelled $1, \cdots, 8$ form the Dynkin diagram of $E_8$, the nodes $0, 1, \cdots, 8$ form the Dynkin diagram of its affine extension $E_8^+$, while the nodes $-1, 0, 1, \cdots, 8$ define its overextension $E_8^{++} \equiv E_{10}$.}
\end{figure}
Both $A_1^{++}$ and $E_{10}$ are hyperbolic, which means that if one removes any node from their Dynkin diagram, one obtains a Dynkin diagram which is either of finite or affine type.  For instance, in the case of $E_{10}$, one gets successively (removing the node $-1$, $0$ etc): $E_8^+$ which is affine, and $A_1 \oplus E_8$, $A_2 \oplus E_7$, $A_3 \oplus E_6$, $A_4 \oplus D_5$, $A_5 \oplus A_4$, $A_6 \oplus A_1 \oplus A_2$, $A_8 \oplus A_1$, $D_9$, $A_9$, wich are all of finite type.

\item The algebra $B_8^{++} \equiv BE_ {10}$ relevant to $N=1$ ten-dimensional supergravity with one vector multiplet.
\begin{figure}[h]
\begin{center}
\includegraphics[scale=.55]{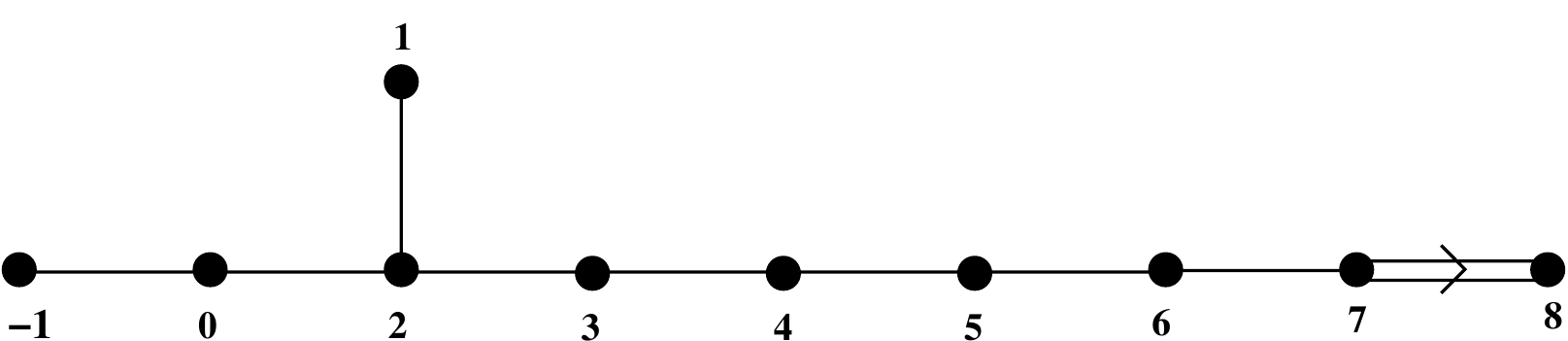}
\end{center}
\caption{The hyperbolic algebra $BE_{10}$. The nodes labelled $1, \cdots, 8$ form the Dynkin diagram of $B_8$, the nodes $0, 1, \cdots, 8$ form the Dynkin diagram of its affine extension $B_8^+$, while the nodes $-1, 0, 1, \cdots, 8$ define its overextension $B_8^{++} \equiv BE_{10}$.}
\end{figure}
As shown in \cite{Henneaux:2003kk}, ``split symmetry controls chaos" and hence, it is the same billiard that controls the dynamics of ten-dimensional supergravity with k vector multiplets, in which case the symmetry algebra in 3 dimensions $so(8, 8+k)$ whose maximal split subalgebra $so(8,9)$.  The algebra $BE_{10}$ is easily verified to be hyperbolic.

\item The algebra $D_8^{++} \equiv DE_ {10}$ relevant to pure $N=1$ ten-dimensional (Fig 4).
\begin{figure}[h]
\begin{center}
\includegraphics[scale=.55]{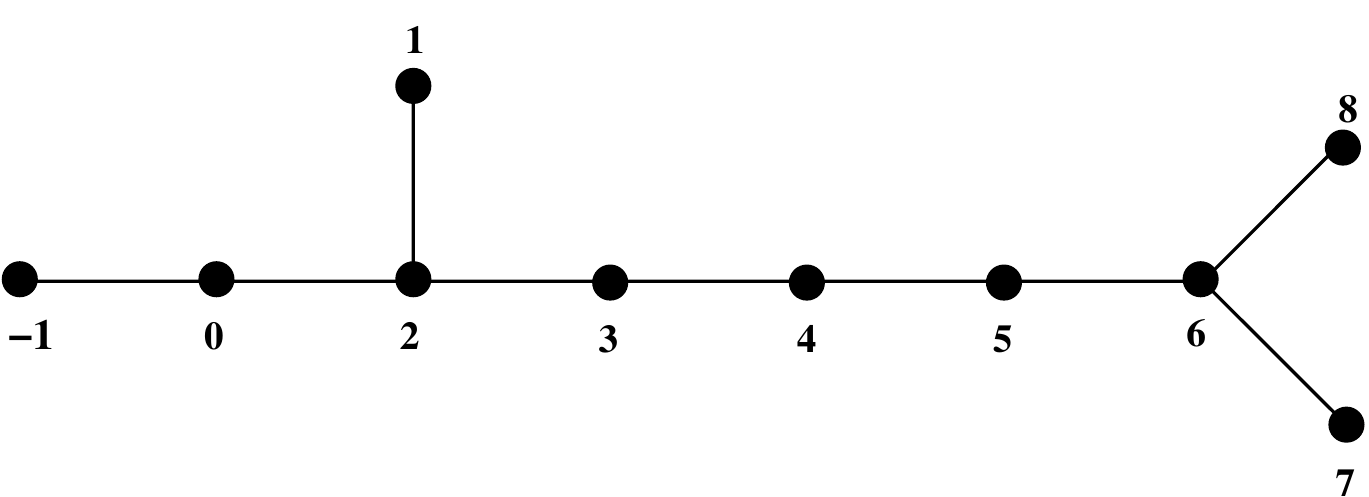}
\end{center}
\caption{The hyperbolic algebra $DE_{10}$. The nodes labelled $1, \cdots, 8$ form the Dynkin diagram of $D_8$, the nodes $0, 1, \cdots, 8$ form the Dynkin diagram of its affine extension $D_8^+$, while the nodes $-1, 0, 1, \cdots, 8$ define its overextension $D_8^{++} \equiv DE_{10}$.}
\end{figure}
The algebra $DE_{10}$ is also easily verified to be hyperbolic.
\end{itemize}

\break

We stress that the emergence of the Kac-Moody structure holds for all theories whose toroidal dimensional reduction to three dimensions has the properties indicated above, but it is by no means necessary to actually perform the reduction to three dimensions to get the billiard.  This follows from the dynamics for {\em generic} initial conditions.

It also turns out that the billiard table is a fundamental domain for the action of the Weyl group on hyperbolic space (upper sheet of the unit hyperboloid).  That it is acute-angled is crucial in this respect.

The fact that one gets the Weyl group of a Lorentzian Kac-Moody algebra is rather remarkable as it depends on the presence of all walls.  As analysed in \cite{Henneaux:2008me} removing some walls, as dictated for instance by spatial cohomology in a regime of intermediate asymptotics \cite{Wesley:2005bd}, might yield Coxeter groups that are not equal to Weyl groups. One might then get a billiard table that is not acute-angled, or which is not a simplex\footnote{It should also be recalled that in hyperbolic space, Coxeter polyhedra need not be simplices.}. Getting the Weyl group of a Kac-Moody algebra is thus a quite non trivial phenomenon.

{}Finally, we note that hyperbolic Kac-Moody algebras exist only up to rank $10$.  In rank $10$, there are four of them, namely, $E_{10}$, $BE_{10}$, $DE_{10}$ already encountered, as well as $CE_{10}$ which is the algebra dual to $BE_{10}$ (its Dynkin diagram is obtained by reversing the arrow connecting the nodes 7 and 8).  The Weyl group of rank 10 algebras act naturally on $9$-dimensional hyperbolic space, and hyperbolicity translates itself in the property that the billiard table (fundamental domain) has finite volume.  The disappearance of chaos for pure gravity as one increases the spacetime dimensions follows from the fact that the algebras $A_k^{++}$ are hyperbolic up to $A_7^{++}$, while $A_k^{++}$ is not hyperbolic for $k \geq 8$ \cite{Damour:2001sa}.

For information we note that while simplex Coxeter groups exist in hyperbolic space up to dimension 9 as we have just recalled, non simplex Coxeter groups exist in hyperbolic space up to dimension $996$ \cite{Vinberg}.

\section{Motion in Cartan subalgebra}
One can precisely reformulate the billiard dynamics as a motion in the Cartan subalgebra of the associated Kac-Moody algebra.  To that end, let us recall the basic features of a Kac-Moody algebra \cite{Kac}.

A Kac-Moody algebra is defined by a (generalized) Cartan matrix $A_{ij}$ ($i, j = 1, \cdots, N$), namely, a (square) matrix with the following properties:
\begin{itemize}
\item  $A_{ii} = 2$
\item  $A_{ij} \in Z_-$ ($i \not=j$)
\item  $A_{ij}\not=0 \Rightarrow A_{ji} \not= 0.$
\end{itemize}
The corresponding Kac-Moody algebra $\mathcal{A}$ is generated by $3N$ generators $\{h_i, e_i, f_i\}$ ($i = 1, \cdots, N=r+2$) subject to the following relations
\begin{eqnarray} && [h_i,h_j] = 0 \label{HH} \\ &&[h_i, e_j] = A_{ij} e_j, \; \; [h_i, f_j] = -A_{ij} f_j , \; \; [e_i, f_j] = \delta_{ij} h_i \label{Chev}\\
&& \underbrace{[e_i, [e_i, [e_i, [ \cdots, [e_i, e_j]] \cdots ]}_{1-A_{ij} \hbox{ times}} = 0 , \; \; \underbrace{[f_i, [f_i, [f_i, [ \cdots, [f_i, f_j]] \cdots ]}_{1-A_{ij} \hbox{ times}} = 0 \; \; \; \; \; \; \; \label{Serre} \end{eqnarray}
Relations (\ref{HH}) and (\ref{Chev}) are the Chevalley relations, relations (\ref{Serre}) are the Serre relations.
The integer $N$ is called the {\em rank} of the algebra.

A central feature of Kac-Moody algebra is the triangular decomposition,
$$ \mathcal{A} = \mathcal{N}^- \oplus H  \oplus \mathcal{N}^+ $$
where (i) $H$ is the Cartan subalgebra (linear combinations of $h_i$); (ii)
$\mathcal{N}^+$ contains the linear combinations of the ``raising operators" $e_i$ and their multiple commutators $[e_i, e_j]$, $[e_i, [e_j, e_k]]$ etc not killed by the above relations; and (iii) $\mathcal{N}^-$ contains the linear combinations of the ``lowering operators" $f_i$ and their multiple commutators $[f_i, f_j]$, $[f_i, [f_j, f_k]]$ etc not killed by the above relations.  This generalizes the well-known triangular decomposition of finite-dimensional simple Lie algebras, e.g.
$$ \left(
     \begin{array}{ccc}
       a & b & c \\
       d & e & f \\
       g & h & -a -e \\
     \end{array}
   \right) = \left(
               \begin{array}{ccc}
                 0 & 0 & 0 \\
                 d & 0 & 0 \\
                 g & h & 0 \\
               \end{array}
             \right)
             + \left(
                 \begin{array}{ccc}
                   a & 0 & 0 \\
                   0 & e & 0 \\
                   0 & 0 & -a-e \\
                 \end{array}
               \right) + \left(
                           \begin{array}{ccc}
                             0 & b & c \\
                             0 & 0 & f \\
                             0 & 0 & 0 \\
                           \end{array}
                         \right)
$$
for $sl(3)$.

One has $$ [h, e_i] = \alpha_i(h) \, e_i $$ where $\alpha_i \in H^\star $ are the {\em simple roots}.
If  $$[e_{i_1}, [e_{i_2}, [\cdots, [e_{i_{m-1}}, e_{i_m}]] \cdots ]] \not= 0, $$
then $\alpha_{i_1} + \alpha_{i_2} + \cdots \alpha_{i_m}$ is a (positive) root,
\begin{eqnarray}&& [h, [e_{i_1}, [e_{i_2}, [\cdots, [e_{i_{m-1}}, e_{i_m}]] \cdots ]]] \nonumber \\ && \hspace{.5cm} = (\alpha_{i_1} + \alpha_{i_2} + \cdots \alpha_{i_m})(h) \, \, [e_{i_1}, [e_{i_2}, [\cdots, [e_{i_{m-1}}, e_{i_m}]] \cdots ]]\nonumber \end{eqnarray} (Jacobi identity).  One has similar relations on the negative side.
If the matrix $A_{ij}$ is symmetrizable, $$A_{ij} = 2 d_i \, S_{ij}, \; \; \; d_i > 0, \; \; \; S_{ij}= S_{ji},$$ as we shall assume here, on can define a scalar product in the real linear span of the simple roots, $$(\alpha_i \vert \alpha_j) = S_{ij}.$$  It is customary to normalize the scalar product such that the longest roots have length squared equal to 2.

One distinguishes 3 cases, according to which the scalar product is Euclidean (``finite case"), positive semi-definite (``affine case") or of neither of these two types (``indefinite case").  It is only in the Euclidean case that the algebra is finite-dimensional.  In the other two cases, it is infinite-dimensional.

In the affine and indefinite cases, a root can be real (= spacelike) or imaginary (= timelike or null).
Simple roots are real; real roots are similar to roots of finite-dimensional, simple Lie algebras: they are non-degenerate and furthermore, if $\alpha$ is a real root, the only multiples of $\alpha$ that are roots are $\pm \alpha$.
By contrast, the imaginary roots do not enjoy these properties and are poorly understood.  The indefinite case with Lorentzian signature is called ``Lorentzian".

The Weyl group of $\mathcal{A}$ is generated by {\em fundamental Weyl reflections}: $$ s_i: \; \lambda \rightarrow s_i(\lambda) = \lambda - 2 \frac{(\lambda\vert \alpha_i)}{(\alpha_i \vert \alpha_i)} \alpha_i . $$
It is a discrete subgroup of $O(N-1,1)^+$ in the Lorentzian case (time orientation preserving elements of $O(N-1,1)$).  It has a well defined action on the upper light cone and on the upper sheet of the unit hyperboloid ($(N-1)$-dimensional hyperbolic space).
The fundamental Weyl chamber is defined in terms of the simple roots by $\alpha_i(h) \geq 0$.  If it is completely contained within the light cone, the algebra is called hyperbolic.  Its intersection with hyperbolic space has then finite volume and is a fundamental domain for the action of the Weyl group on hyperbolic space.

The dictionary between the billiard motion and the Kac-Moody algebra is given in the following table \cite{DH2,Damour:2001sa}
\vspace{.3cm}

\begin{tabular}{|c| c |c|}
  \hline
  {\bf Gravity side}  &  & {\bf Kac-Moody side} \\
  \hline
  \hline
  Scale factors  & $\leftrightarrow$ & Cartan degrees of freedom \\
  \hline
  Billiard motion  & $\leftrightarrow$ & Lightlike motion  \\
   &  & in Cartan subalgebra \\
  \hline
  Walls & $\leftrightarrow$ & Hyperplanes orthogonal  \\
   &  & to simple roots \\
   \hline
  Reflection against a wall & $\leftrightarrow$ & Fundamental Weyl reflection \\
  \hline
  Finite volume of billiard table & $\leftrightarrow$ & Hyperbolic algebra \\
  \hline
\end{tabular}

\section{Hidden Symmetries?}
The intriguing emergence of the Weyl group of a Kac-Moody algebra in the BKL limit has prompted the conjecture that the Kac-Moody algebra itself might be a hidden symmetry of the corresponding gravitational theory (possibly augmented by new degrees of freedom)\cite{Damour:2002cu}.  Part of the excitement regarding this conjecture is due to the fact that the same conjecture was made (earlier in the case of some of the approaches) following different lines in \cite{Julia,Julia2,Nicolai:1991kx,West1,HenryLabordere:2002dk,Englert1}.  Attempts to substantiate the conjecture have been made based on the idea of non linear realizations, in which the conjectured symmetry is manifest \cite{West1,Damour:2002cu}. The problem is then to connect the dynamics of the non linear sigma model to the (super)gravity dynamics through an appropriate dictionary and to establish their equivalence (with the possible addition of new degrees of freedom on the gravity side).  It is not the purpose here to review all the interesting work that has gone into studying various aspects of this conjecture.  We shall only allude to the approach inspired by the cosmological billiards, which is the exclusive subject of this article.

The BKL analysis suggests to consider the ($1+0$)-non linear sigma model $G^{++}/K(G^{++})$ (geodesic motion on $G^{++}/K(G^{++})$) \cite{Damour:2002cu}. Namely, one goes beyond the dynamics in the Cartan subalgebra by including as dynamical variables the fields associated with the positive roots.  This approach has met with spectacular successes at low ``levels" \cite{Damour:2002cu} (see \cite{Damour:2004zy} for a systematic analysis) since its low level truncation reproduces the dynamics of supergravity consistently truncated to homogeneous modes (in a sense made precise in \cite{Damour:2002cu}).  But no one has been able to push it systematically to higher levels so far to include the full supergravity theory.  Further work (and new ideas) appears to be necessary.  Part of the problem is that the dictionary between the sigma model variables and the supergravity fields is not understood. A better control of duality appears also to be necessary. The same success works, up to the same levels, if one includes the fermionic degrees of freedom, leading, on the sigma-model side, to a spinning particle action \cite{deBuyl:2005zy,Damour:2005zs,de Buyl:2005mt,Damour:2006xu}

Two other spectacular findings provide additional support to the conjecture:
\begin{itemize}
\item The quantum corrections to $M$-theory are compatible with the $E_{10}$ algebraic structure, in the sense that they correspond to roots of $E_{10}$ \cite{Damour:2005zb} (see also \cite{Lambert:2006he,Damour:2006ez,Lambert:2006ny,Bao:2007er,Bao:2007fx} for further discussions and developments).
\item The massive deformation of type IIA supergravity corresponds to a level 4 root of $E_{10}$ (or $E_{11}$) \cite{Schnakenburg:2002xx,deBuyl}, a result that can be extended to other deformations of the theory in lower dimensions (see the original works \cite{Bergshoeff:2007qi,Riccioni:2007au,West2,Bergshoeff} for entries into this fast growing literature).
\end{itemize}

\section{Conclusions}
We have reviewed the dynamics of the gravitational field in the vicinity of a cosmological singularity pioneered by the remarkable work of BKL and have shown that it can be described in terms of fascinating structures: hyperbolic Coxeter groups and Kac-Moody algebras.  This seems to be the tip of an iceberg indicating an even more richer structure at a deeper level, yet to be discovered.

\section*{Acknowledgements}
It is a pleasure to thank the various people with whom I had the priviledge to work on this problem over the years, and in particular the late Jacques Demaret with whom the analysis of BKL in higher dimensions was started, and Thibault Damour, with whom the BKL study was given a new impetus leading to results (connection with Coxeter groups and Kac-Moody algebras), the significance of which are still being explored.  Work supported in part by IISN-Belgium
(conventions 4.4511.06 and 4.4514.08), by the Belgian National Lottery, by the European
Commission FP6 RTN programme MRTN-CT-2004-005104, and by the Belgian Federal Science Policy Office through the
Interuniversity Attraction Pole P6/11.

\end{document}